\documentclass[%
reprint,
superscriptaddress,
showpacs,preprintnumbers,
 amsmath,amssymb,
 aps,
prb,
]{revtex4-1}

\usepackage{graphicx}
\usepackage{dcolumn}
\usepackage{bm}
\usepackage[version=3]{mhchem}
\usepackage{miller}

\begin{document}


\title{Complex oxide growth using simultaneous \textit{in situ} RHEED and x-ray reflectivity:\\When is one layer complete?}

\author{M.\ C.\ Sullivan} \email{mcsullivan@ithaca.edu}
\affiliation{Department of Physics and Astronomy, Ithaca College, Ithaca NY, USA}
\affiliation{School of Applied and Engineering Physics, Cornell University, Ithaca NY, USA}

\author{M.\ J.\ Ward}
\affiliation{Cornell High Energy Synchrotron Source, Cornell University, Ithaca NY, USA}

\author{Araceli Guti\'{e}rrez-Llorente}
\affiliation{School of Applied and Engineering Physics, Cornell University, Ithaca NY, USA}

\author{Eli R.\ Adler}
\affiliation{Department of Physics and Astronomy, Ithaca College, Ithaca NY, USA}
\affiliation{Cornell High Energy Synchrotron Source, Cornell University, Ithaca NY, USA}

\author{H. Joress}
\affiliation{Cornell High Energy Synchrotron Source, Cornell University, Ithaca NY, USA}
\affiliation{Department of Materials Science and Engineering, Cornell University, Ithaca NY, USA}

\author{A.\ Woll}
\affiliation{Cornell High Energy Synchrotron Source, Cornell University, Ithaca NY, USA}

\author{J.\ D.\ Brock}
\affiliation{School of Applied and Engineering Physics, Cornell University, Ithaca NY, USA}
\affiliation{Cornell High Energy Synchrotron Source, Cornell University, Ithaca NY, USA}

\date{\today}

\begin{abstract}
During layer-by-layer homoepitaxial growth, both the Reflection High-Energy Electron Diffraction (RHEED) intensity and the x-ray reflection intensity will oscillate, and each complete oscillation indicates the addition of one monolayer of material.  However, it is well documented, but not well understood, that the phase of the RHEED oscillations is not constant and thus the maxima in the RHEED intensity oscillations do not necessarily occur at the completion of a layer.  We demonstrate this using simultaneous \textit{in situ} x-ray reflectivity and RHEED during layer-by-layer growth of SrTiO$_3$.  We show that we can control the RHEED oscillation phase by changing the pre-growth substrate annealing conditions, changing the RHEED oscillation phase by nearly 180$^\circ$.  In addition, during growth via pulsed laser deposition, the exponential relaxation times between each laser pulse can be used to determine when a layer is complete, independent of the phase of the RHEED oscillation.
\end{abstract}

\maketitle



Thin-film growth changed dramatically more than three decades ago with the discovery of reflection high-energy electron diffraction (RHEED) intensity oscillations.\cite{Harris1981, Harris1981a, Wood1981}  During RHEED, a high-energy ($\approx$ 10-30 keV) electron beam is fired at grazing incidence onto a growth surface and the intensity of the reflected beam is recorded.  With RHEED, the incident electrons only interact with the topmost layer.  During 2D monolayer-by-monolayer growth, researchers discovered oscillations in the intensity of the reflected RHEED beam, and that the period of the oscillation corresponded to the addition of exactly one monolayer to the film.  This discovery led to rapid implementation of RHEED systems for thin-film growth, although largely restricted to semiconductor growth via molecular beam epitaxy due to the low pressures ($<10^{-5}$ mbar) required to use RHEED.\cite{Neave1983,Snyder1991}

The landscape changed again nearly two decades ago with the development of high-pressure RHEED systems, allowing RHEED systems to operate at pressures as high as 1 mbar.\cite{Rijnders1997, Blank1998, Blank1999} Since this discovery, \textit{in situ} RHEED characterization has become nearly ubiquitous in thin-film growth systems, and it has been successfully implemented in a variety of growth techniques in addition to molecular beam epitaxy, such as sputtering\cite{shih1994} and pulsed laser deposition (PLD).\cite{Koster2011}

Researchers have worked to understand RHEED intensity oscillations via a variety of different methods.  In principle, the complete picture can only be understood using dynamical diffraction theories,\cite{braun1998, Mitura1998} which often have to be modified for complicated growth conditions (e.g., including  variations in the scattering potential\cite{mitura1999} or small terrace sizes\cite{mitura2002}). Before resorting to a full model, it often suffices to describe the RHEED intensity oscillations as the interference between two layers via a kinematic scattering approximation,\cite{lent1984,pukite1985} though this model can become more complex when multiple layers are included\cite{kawamura1985} or other diffraction features such as Kikuchi lines are included.\cite{shin2007a}  The other often-used simplification is the step density model,\cite{Neave1983, Shitara1992} which predicts the decrease of the reflected specular RHEED intensity as the areal density of steps increases.  It is also common practice to use a combination of these models.\cite{shin2007b, Koster2011}

Many of the models, especially the simpler step density model and the kinematic approximation, share similar traits.  Chief among those traits is the prediction that the specular RHEED intensity decreases as the surface becomes rougher.  For example, in the step density model, as the step density increases the likelihood of diffuse scattering increases, thus the specular intensity decreases.  As the layer reaches completion, the number of steps decreases, and thus the specular intensity recovers -- ideally returning to its original value.\cite{Koster2011}
The prediction is the same for the kinematic approximation: we expect the RHEED intensity to oscillate exactly out of phase with the surface roughness.

In practice, RHEED oscillations are often more complex: the RHEED intensity can \textit{increase} at the start of the growth,\cite{Haeni2000} or decrease and then recover to an intensity \textit{greater} than the intensity before growth,\cite{lippmaa2000, ohtomo2007} or the RHEED intensity can (of course) oscillate exactly out of phase with the roughness.\cite{Koster1998}  Even identical RHEED and growth conditions can yield RHEED oscillations that are surprisingly 180$^\circ$ out-of-phase with previous oscillations.\cite{Haeni2000}

RHEED oscillations during PLD have an added layer of complexity due to the movement of adatoms between laser pulses.  During PLD, each laser pulse ablates a large amount of material from the target.  This material is deposited randomly on the surface of the substrate, creating a sharp decrease in intensity after each laser pulse.  Between laser pulses, the adatoms diffuse on the surface, falling into pits and attaching to step edges and islands, and thus the surface ``heals'' and becomes smoother.  As the surface heals between laser pulses, the RHEED intensity increases.  This behavior agrees with the simple RHEED oscillation models -- despite the fact that the overall growth oscillation is not necessarily out of phase with the roughness.\cite{lippmaa2000}  This can be a major drawback when using RHEED -- it is not always obvious when a layer is complete.

In order to quantify the RHEED oscillations, researchers have defined the phase of the RHEED intensity oscillations.\footnote{In PLD, these oscillations refer to the large-scale intensity oscillations, not the change in intensity between laser pulses.  Typical RHEED intensity oscillations have a period of 10-50 laser pulses.}  The period $T$ of the oscillations is easy to define based on the maxima or minima of the intensity oscillations.  Following previous work, we assume the growth begins at $t=0$ and look for the minimum of the RHEED intensity that occurs between $t=T$ and $t=2T$ and label this time as $t_{3/2}$.\cite{Zhang1987}  Then the phase of the RHEED oscillations is defined as\cite{Mitura1998}
\begin{equation}
\phi = 2 \pi(t_{3/2}/T - 1.5).\label{eq:t3-2}
\end{equation}
Using the simple models we expect $t_{3/2} = 1.5 T$ (RHEED intensity is a minimum when the surface is roughest), so $\phi = 0$, indicating that the RHEED intensity is in phase with the smoothness of the substrate and exactly out of phase with the roughness.

Despite the proliferation of RHEED systems, it is not the only \textit{in situ} growth diagnostic tool.  With a suitably chosen scattering geometry (typically at incident angles smaller than that of the first Bragg peak), x-ray scattering is highly surface sensitive and can also yield information about growth dynamics.\cite{Fleet2005, Dale2006}  X-ray reflectivity (XRR) probes much deeper than RHEED and contains additional information about layers below the topmost layer.  As a result, XRR intensity at angles below the first Bragg peak will oscillate not only with roughness but also due to thin-film interference.  However, in the case of homoepitaxy, there is no thin-film interference, and therefore the XRR intensity oscillates due to roughness only.  Because x-rays are weakly interacting, the kinematic scattering approximation describes XRR intensity oscillations very accurately.\cite{Woll2011}  As a result, we can use Eq.\ \ref{eq:t3-2} to find $\phi = 0$ always for homoepitaxial XRR intensity oscillations.

In addition, the intensity lost from the XRR specular reflections can be seen directly in the XRR diffuse scattering\cite{Fleet2006, Ferguson2009, Brock2010} (something that is very difficult to capture with RHEED).  In fact, the diffuse scattering can be used to determine the distance between islands on the surface of the substrate.\cite{Ferguson2009, Brock2010}  However, at these angles far from a Bragg peak, nearly complete destructive interference means that very bright synchrotron x-ray sources are required to be able to see XRR intensity oscillations, which is why XRR continues to be less common than RHEED.

In this article we discuss direct, simultaneous comparison of  RHEED and XRR during homoepitaxial growth of \ce{SrTiO3} (STO) via pulsed laser deposition onto STO \hkl<001> substrates.  We used a KrF eximer laser ($\lambda = 248$ nm) 
with a fluence of $\approx2$ J/cm$^2$, 
with a spot size on the target of $3.7$ mm$^2$, 
yielding approximately 10 laser pulses per monolayer.  Depositions occurred at 900 $^\circ C$ 
and an O$_2$ pressure of $1.3\times10^{-5}$ mbar. 
Before deposition, the substrates are etched in HF and annealed to produce an atomically smooth \ce{TiO2} terminated surface.\cite{Koster1998, Kawasaki1994}   Further experimental details can be found in Ref.\ \onlinecite{Ferguson2009}.

The x-ray and RHEED measurements were performed using a custom PLD/x-ray diffraction system installed in the G3 hutch at the Cornell High Energy Synchrotron Source.  X-ray measurements were taken at the ``anti-Bragg'' or ``quarter-Bragg'' positions ( the \hkl(0 0 \frac{1}{2})  or the \hkl(0 0 \frac{1}{4}))  in reciprocal space and images collected using a Pilatus 100K detector.  The RHEED system is attached to the chamber at 45$^\circ$ to the x-ray beam.  For our substrates, the miscut ran parallel to the \hkl(110) direction, usually to within $\pm5^\circ$, so we were able to align the incident x-ray beam perpendicular to the miscut and the RHEED beam along the \hkl(100) axis.  We used the average intensity of the specular reflection for our RHEED oscillations, and the incident angle of the electron beam varied from 0.8$^\circ$ to 1.5$^\circ$.

\begin{figure}
\includegraphics[width=\linewidth,clip=true]{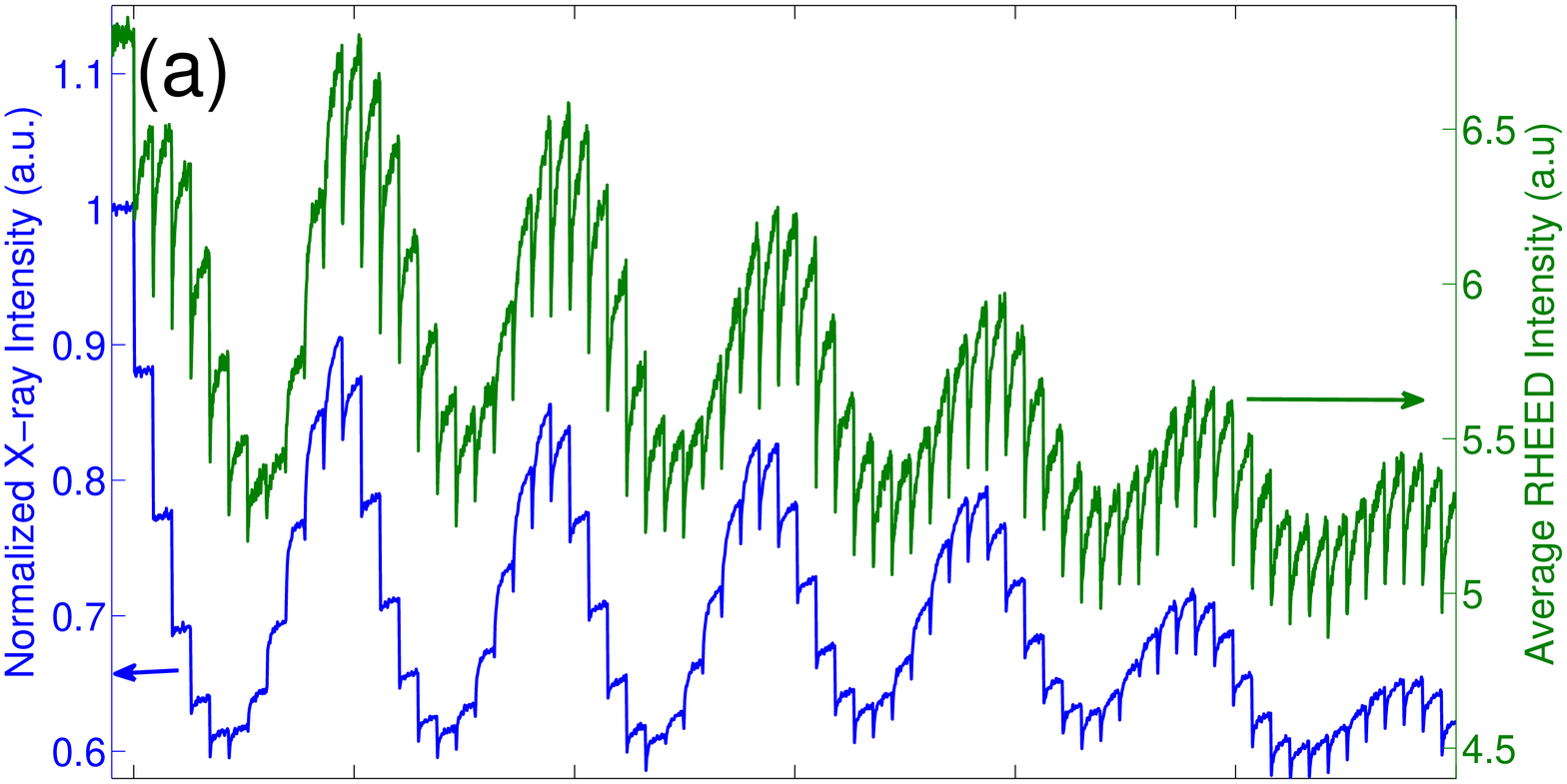}
\includegraphics[width=\linewidth,clip=true]{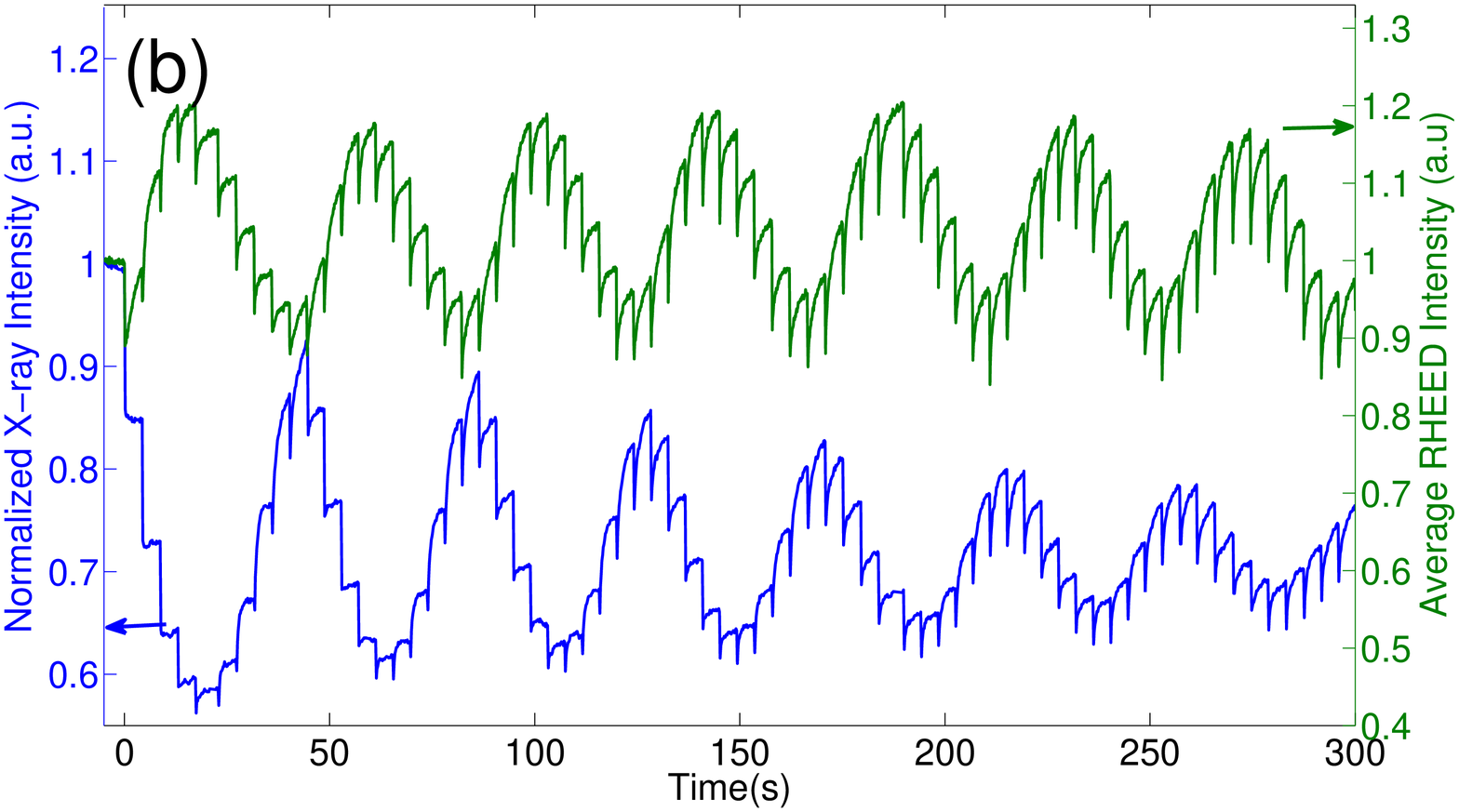}
\caption{(Color online) Simultaneous RHEED and XRR intensity oscillations for two growths.  XRR was measured at the quarter-Bragg position \hkl(0 0 \frac{1}{4})), the RHEED was parallel to the  the \hkl(100) axis.   In both, the RHEED intensity (green) is above the XRR intensity (blue).  The growth conditions were identical.  However, the substrate in (a) was annealed for one hour in $2.7\times10^{-6}$ mbar, the substrate in (b) was annealed for 20 minutes in $1.7\times10^{-3}$ mbar O$_2$.  Annealing in low vacuum created RHEED oscillations very nearly in phase with the XRR oscillations ($\phi \approx 0.05 \pi$); annealing in oxygen created oscillations very nearly out of phase ($\phi \approx 0.81\pi$).} \label{fig:r-x}
\end{figure}

Two typical growths of STO on STO \hkl<001> via pulsed laser deposition are presented in Fig.\ \ref{fig:r-x}.  XRR data was taken at the quater-Bragg position, \hkl(0 0 \frac{1}{4}) and RHEED beam was parallel to the \hkl(100) axis.  Each individual laser pulse is obvious, marked by the sharp decrease in intensity in both RHEED and XRR, followed by the exponential recovery between pulses.  The number of pulses per layer is $\approx 11.3$ and $\approx 10.2$ for Fig.\ \ref{fig:r-x}(a) and (b), respectively.

\begin{figure*}
\begin{minipage}{0.35\linewidth}
\includegraphics[width=\linewidth,clip=true]{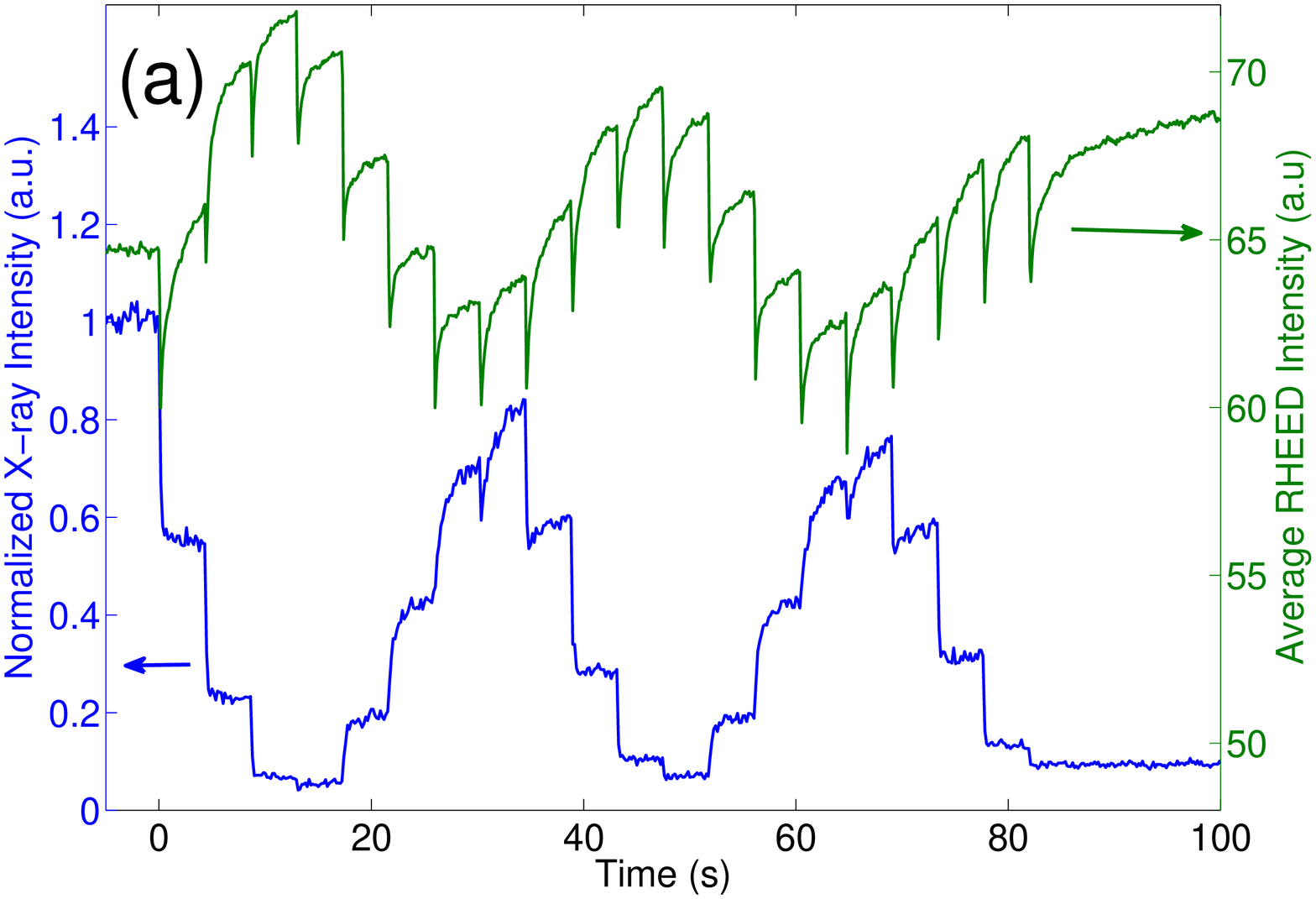}
\end{minipage}
\begin{minipage}{0.25\linewidth}
\includegraphics[width=\linewidth,clip=true]{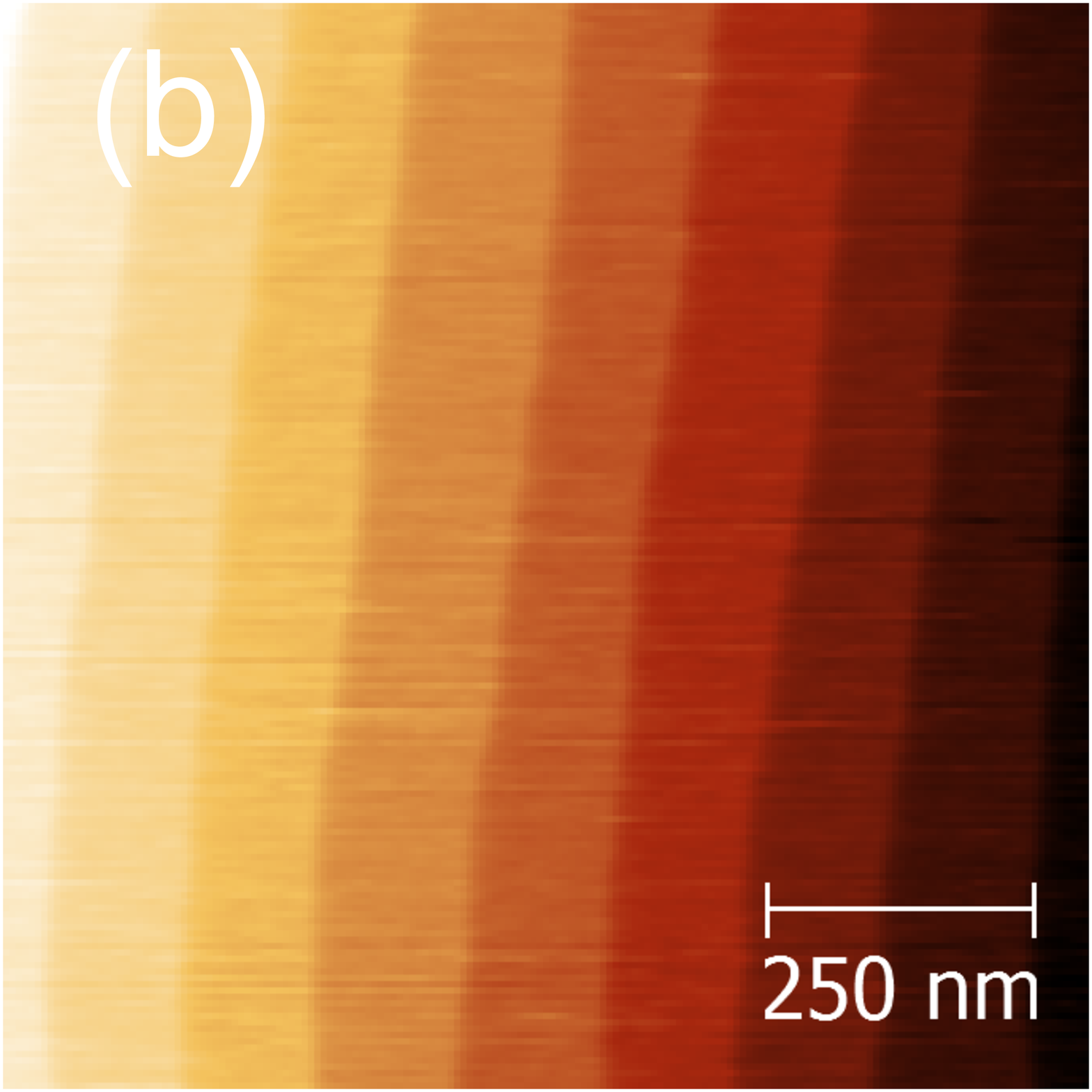}
\end{minipage}
\begin{minipage}{0.25\linewidth}
\includegraphics[width=\linewidth,clip=true]{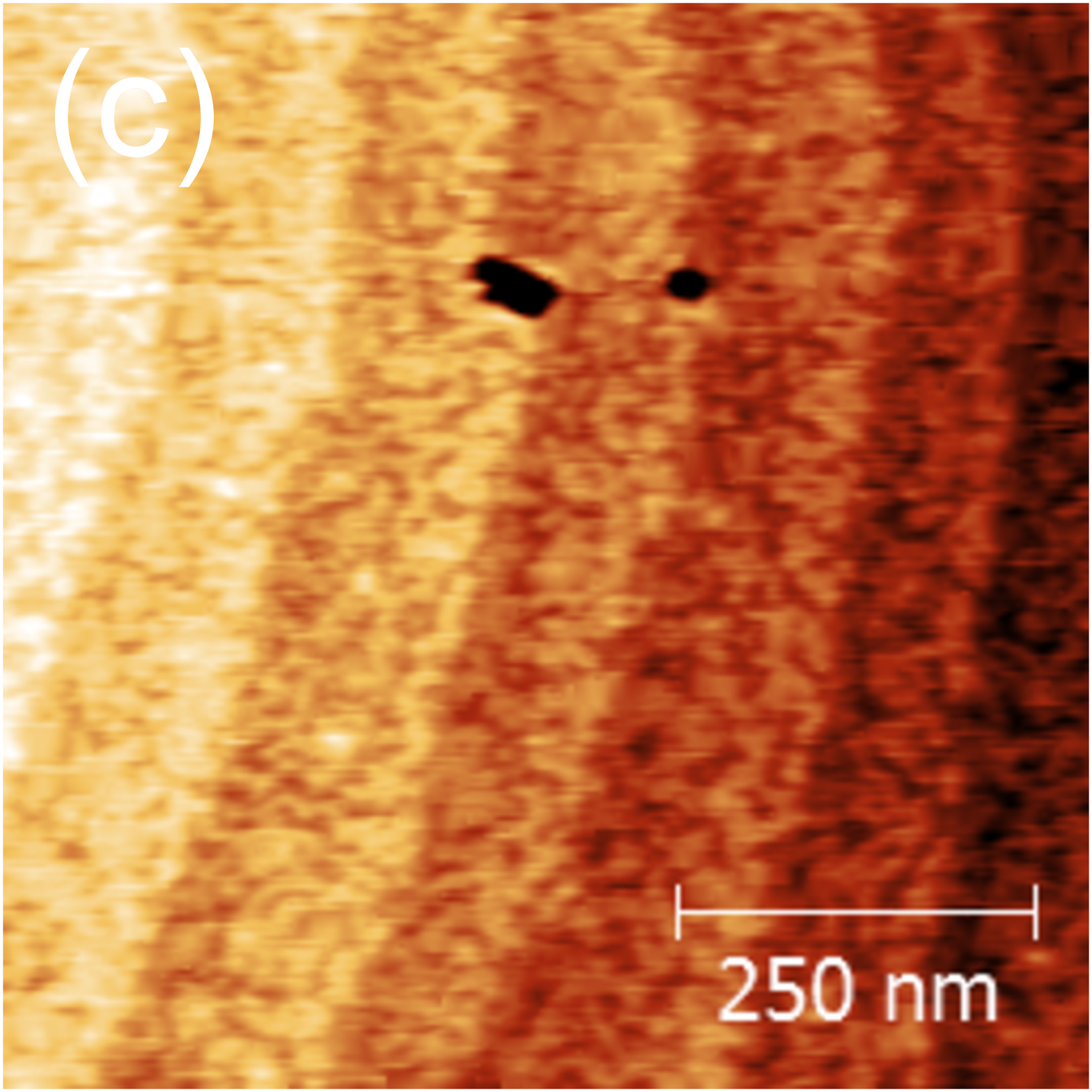}
\end{minipage}
\begin{minipage}{0.35\linewidth}
\includegraphics[width=\linewidth,clip=true]{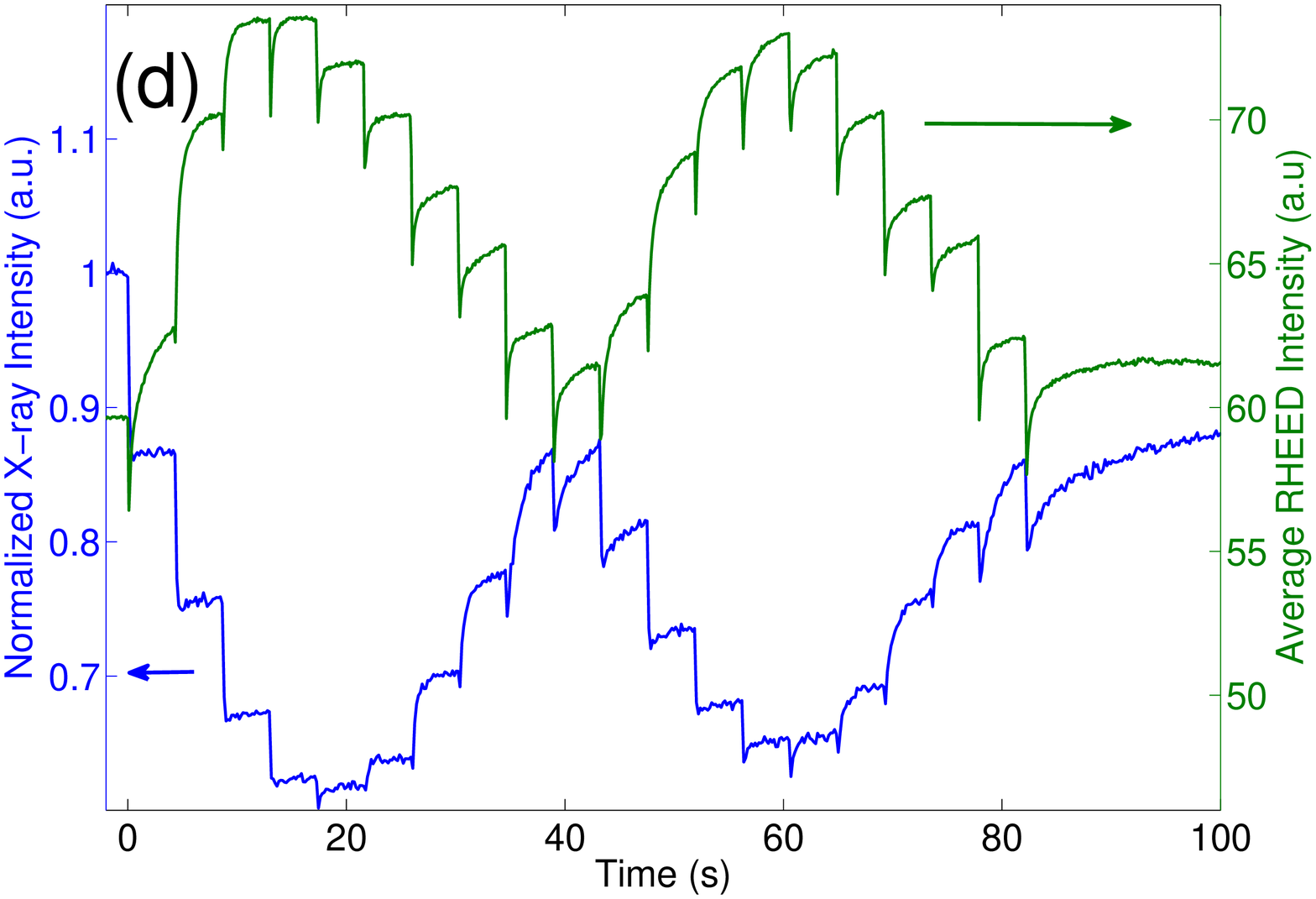}
\end{minipage}
\begin{minipage}{0.25\linewidth}
\includegraphics[width=\linewidth,clip=true]{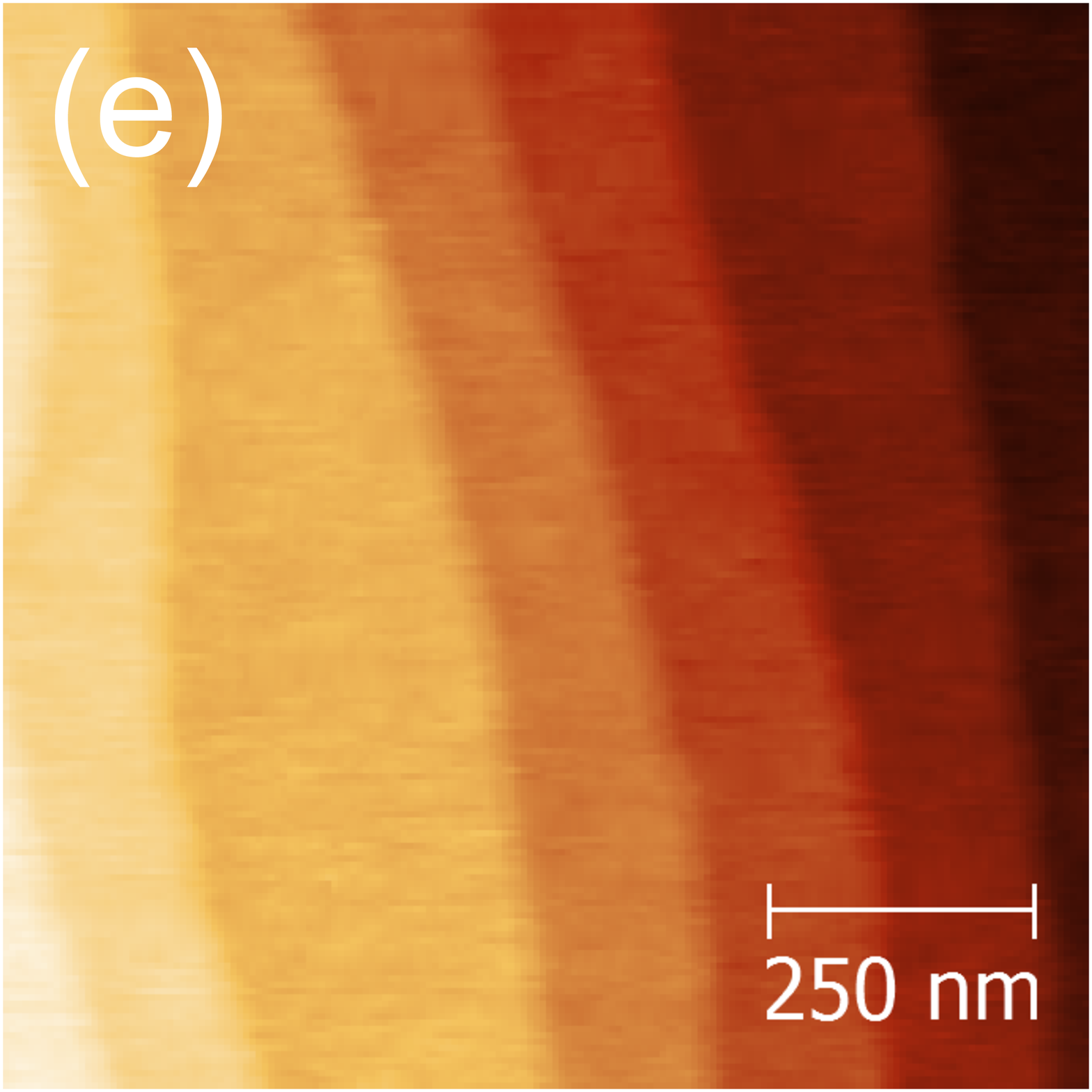}
\end{minipage}
\begin{minipage}{0.25\linewidth}
\includegraphics[width=\linewidth,clip=true]{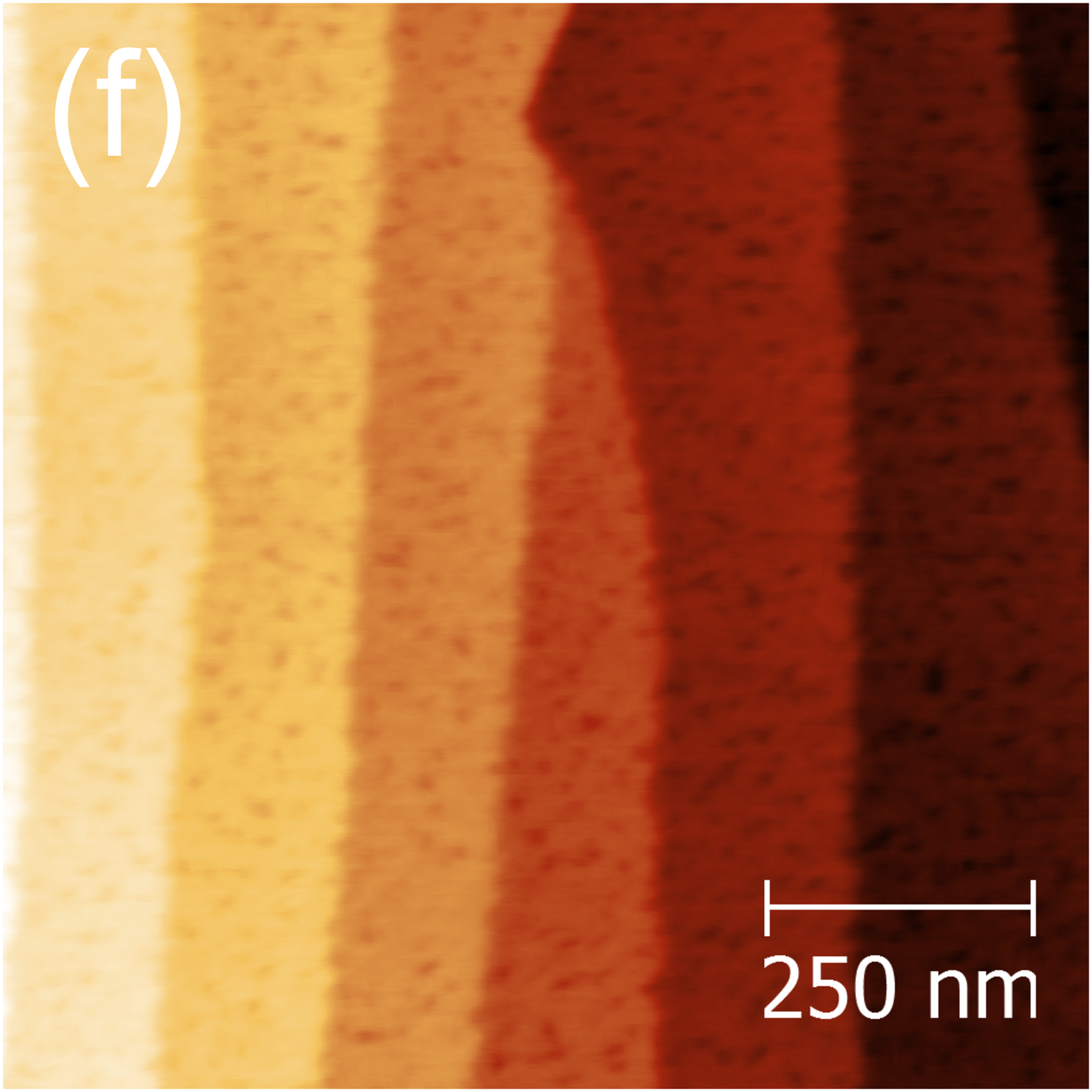}
\end{minipage}
\caption{(Color online) Atomic force microscope images of two growths.  All AFM images are approximately 1 $\mu$m $\times$ 1 $\mu$m.  (a) and (d) show the simultaneous RHEED and XRR intensity oscillations, both nearly exactly out of phase.  XRR is measured along \hkl(0 0 \frac{1}{2})
in (a), along \hkl(0 0 \frac{1}{4}) in (b).  The first growth (a) was interrupted at the minimum of the XRR oscillations and near the maximum of the RHEED oscillations, the second growth (d) was interrupted after three complete monolayers, putting it near the minimum of the RHEED oscillations.  Pre-growth AFM images are shown for both substrates in (b) and (e).  In (c), we see the surface of (a) post growth, where $\approx$50\% of the surface is covered with islands.  The post growth surface of (f) has pinholes over 15\%$\pm$2\% of its surface.\cite{Trofimov2003,Woll2011} For both growths, the maximum of the XRR corresponds to a complete layer, whereas the RHEED intensity cannot be used to determine layer completion.} \label{fig:afms}
\end{figure*}

 The growth conditions of these two films were nearly the same (deposition temperatures of 915 $^\circ C$ and  890 $^\circ C$  for Figs.\ \ref{fig:r-x}(a) and \ref{fig:r-x}(b), respectively).  For both growths, the XRR intensity oscillation remains out of phase with the roughness  (x-ray diffuse scattering), so $\phi = 0$.  However, in  Fig.\ \ref{fig:r-x}(a), the RHEED intensity oscillation is nearly in phase with the XRR oscillation ($\phi \approx 0.05 \pi$), and in Fig.\ \ref{fig:r-x}(b), the RHEED intensity oscillation is nearly out of phase with the XRR oscillation ($\phi \approx 0.81 \pi$).  Other researchers have presented similar growths of STO on STO \hkl<001> that appear to have $\phi \approx -0.62\pi$ (69$^\circ$).\cite{khodan2012}  In our experiments, we can repeatably adjust the phase not via \textit{growth} conditions but rather via substrate  \textit{annealing} conditions.  If the substrate is annealed just before growth in high vacuum, the RHEED oscillations are very nearly in phase; if the substrate is annealed in high \ce{O2} pressure, the oscillations are very nearly out of phase.  The substrate in Fig.\ \ref{fig:r-x}(a) was annealed for one hour in $2.7\times10^{-6}$ mbar, the substrate in Fig.\ \ref{fig:r-x}(b) was annealed for 20 minutes in $1.7\times10^{-3}$ mbar.  Our experiments show that the annealing condition dictates the RHEED oscillation phase for a wide variety of different growth conditions!

Comprehensive studies of the phase of the RHEED oscillations exist only for semiconductors,\cite{Mitura1998} and as such the mechanism that controls the phase of the RHEED oscillation has not been well studied in the oxide materials.  Other researchers have reported results similar to ours for homoepitaxy in STO, with RHEED phases that can vary by 180$^\circ$ for seemingly identical growth conditions.\cite{Haeni2000}  However, to our knowledge, this is the first report of the ability to reliably and repeatably control the RHEED phase.

As a compelling visual confirmation, we characterized our substrates using an atomic force microscope (AFM) before and after growth.  These data are presented in Fig.\ \ref{fig:afms}.   We used the annealing conditions that gave us nearly out-of-phase RHEED and XRR intensity oscillations (in (a), $\phi \approx 0.76\pi$, in (b), $\phi \approx 0.81\pi$).  We interrupted the growth of (a) at $t \approx 2.5T$, or at 2.5 monolayers according to XRR intensity oscillations, which means the RHEED intensity oscillation was close to its maximum.  We can see in Fig.\ \ref{fig:afms}(b) that the substrate was atomically smooth prior to growth, and after growth the sample has a uniform roughness, with islands covering $\approx$ 50\% of its surface.  This maximum roughness occurs at the minimum of the XRR oscillations and thus the maximum roughness, but occurs near the \textit{maximum} of the RHEED oscillation -- providing the clear visual proof that  RHEED can be used to track number of layers but with an uncertainty of $\pm$ half a layer.

Fig.\ \ref{fig:afms}(c) shows a similar growth interrupted at $t\approx 2T$, or 2 monolayers according to the XRR intensity oscillations.   After two monolayers, the atomically smooth substrate before growth (Fig.\ \ref{fig:afms}(e)) has pinholes covering 15\%$\pm$2\% of its surface.\footnote{To determine the coverage in Fig.\ \ref{fig:afms}(f), we masked any area more than 2 \AA~below the level of the terrace and compared the masked area to the total area.  We measured this ratio for two $\approx100$ nm$^2$ areas per terrace. }  These pinholes are expected for layer-by-layer growth and are why the XRR intensity does not recover to its initial maximum.\cite{Trofimov2003,Woll2011}  Here the RHEED intensity is close to its minimum, yet the surface is very smooth.

Most growth systems do not have the capability to measure both RHEED and XRR simultaneously, so the question is: using only RHEED, is there any way to know when a layer is complete?  The answer is simple: yes.  As discussed above, determining the growth period $T$ is straightforward.  If it is true 2D growth and the starting surface is smooth, then each new monolayer is complete at times $t= T, 2T, 3T,\ldots$.  This is true independent of RHEED oscillation phase, and can be seen in Figs.\ \ref{fig:r-x}(a) and (b) as well as Fig.\ \ref{fig:afms}(a) and (d).

For 2D growth, it is still possible to determine when a layer is complete using RHEED -- even if the starting surface is not smooth.  Returning our attention to the intensity as measured between pulses, each laser pulse causes a sharp decrease in intensity followed by an exponential recovery as the adatoms move on the surface and the surface heals.  The step density model predicts a recovery of the form\cite{Blank1999}
 \begin{equation}
   I \approx I_o(1-e^{(t-t_{\mathrm{pulse}})/\tau}), \label{eq:relax}
 \end{equation}
 where $\tau$ is the relaxation time.  When the surface is rough, the relaxation time $\tau$ is short, as it takes very little time for the adatoms to diffuse the short distance required to find a hole, step edge, or island.  On the other hand, when the sample is very smooth, this relaxation time is long, as it takes adatoms a long time before finding one of the few islands or holes on the surface.  

 An increase in relaxation time between laser pulses indicating the completion of a layer has been seen explicitly in XRR.\cite{Fleet2006,Ferguson2009}  Increasing relaxation times as layers reach completion has also been seen in RHEED when the phase $\phi = 0$.\cite{Blank1999} 
 Though it is counterintuitive, the recovery between pulses can be fit using the step density model even when the overall RHEED oscillation is more complex than this simple model.\cite{Blank1998}  Thus, we can use the relaxation time after each laser pulse to characterize the layer coverage, as the relaxation times per pulse will still reach a maximum when the layer is complete, independent of the RHEED oscillation phase.

\begin{figure}
\includegraphics[width=\linewidth,clip=true]{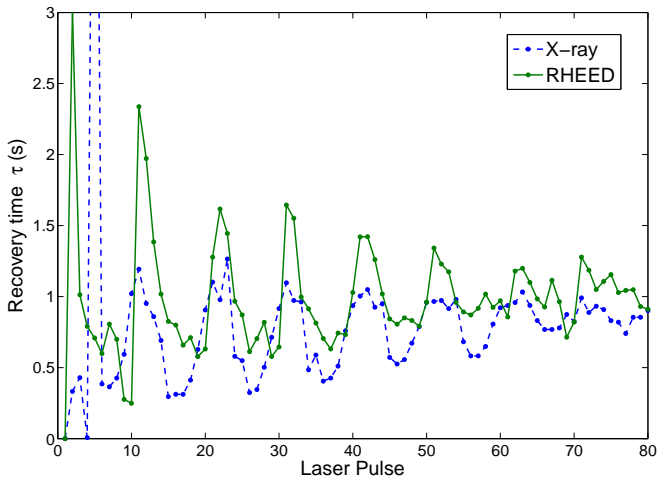}
\caption{(Color online) Recovery times per laser pulse for XRR (blue dashed) and RHEED (green solid).  Error bars (not shown for clarity) are on average $\pm5$\%.  These relaxation times are from the growth in Fig.\ \ref{fig:r-x}(b).   Clear oscillations in the relaxation times can be seen, with maxima occurring approximately every 11 laser pulses.   Here, the maxima in the relaxation times from RHEED and x-ray occur after the same number of laser pulses and occur at the completion of the layer, thus a maximum in the RHEED relaxation time signals the completion of a layer.} \label{fig:relax}
\end{figure}

Using Eq.\ \ref{eq:relax} we fit the recovery after each laser pulse for the growth shown in Fig.\ \ref{fig:r-x}(b), where $\phi \approx 0.81\pi$.  In Fig.\ \ref{fig:relax}, the blue dashed curve represents the relaxation times per pulse as measured by XRR and the green solid line represents the relaxation times as measured by RHEED.  As expected, the XRR relaxation times are a maxima when the layer is complete, roughly every 11 laser pulses.  Since the XRR intensity is at its maximum when the layer is complete, the oscillations of the XRR relaxation times are roughly in phase with the XRR intensity oscillations.  Remarkably, the maxima in the relaxation times in the RHEED data  occur at the same laser pulse as the XRR relaxation times, that is to say the  relaxation times measured by both techniques are a maxima at $t = T, 2T, 3T,\ldots$. Thus, the RHEED relaxation times are a maximum when the layer is complete -- despite the fact that the RHEED intensity oscillation phase is nearly 180$^\circ$!

Similar behavior has been seen at other RHEED intensity oscillation phases.  Khodan \textit{et al.} grew STO on STO \hkl<0 0 1> via PLD\cite{khodan2012}   and used substrates etched using a similar HF etch.  Assuming a smooth starting surface, their data present an oscillation phase of $\phi \approx -0.62\pi$ (69$^\circ$),\cite{khodan2012} with RHEED relaxation times that are a maxima at $t= T, 2T, 3T\ldots$, again as expected.

In conclusion, we have studied the homoepitaxial growth of STO on STO \hkl<001> via simultaneous \textit{in situ} RHEED and XRR. We have shown that the RHEED intensity oscillation phase $\phi$ can change even for identical growth conditions, and that in contrast the XRR intensity oscillations are always at a maximum when the layer is complete ($\phi = 0$ for XRR).  From post-growth AFM images, we have shown that the substrate surface can be rough even when the RHEED intensity oscillation is near a maximum and smooth when the RHEED oscillation is near a minimum.

Finally, the main point of this article is to provide a tool to the oxide growth community to determine when a layer is complete, a tool that does not depend on the magnitude of the RHEED intensity oscillation.  For PLD, the RHEED and XRR intensities increase after each laser pulse as the adatoms diffuse and the surface heals.  The characteristic relaxation time between each laser pulse is a maximum when the surface is least rough, and can be used to determine when a layer is complete.  We have shown in our own results that this relaxation time is a maximum at layer completion for various phases of RHEED growth.

\begin{acknowledgments}
The authors  acknowledge  Hanjong  Paik and Charles Brooks for  helpful  discussions and assistance in etching substrates and  Darrell Schlom  for  use of his laboratory facilities for etching some substrates.  The remainder of our substrates were etched at the Cornell NanoScale Facility, a member of the National Nanotechnology Infrastructure Network, which is supported by the National Science Foundation (Grant ECCS-0335765).

M.\ C.\ Sullivan was supported in part by the Energy Materials Center at Cornell (EMC2), an Energy Frontier Research Center funded by the U.S. Department of Energy, Office of Science, Office of Basic Energy Sciences under Award Number DE-SC0001086.

This work is based upon experiments conducted at the Cornell High Energy Synchrotron Source (CHESS) which is supported by the National Science Foundation and the National Institutes of Health/National Institute of General Medical Sciences under NSF awards DMR-1332208 and DMR-0936384.  This work also made use of the Cornell Center for Materials Research Shared Facilities which are supported through the NSF MRSEC program (DMR-1120296).
\end{acknowledgments}

\nocite{*}

\bibliography{r_x_refs}

\end{document}